\documentclass[a4paper,10pt,twocolumn,amsmath,amssymb]{article}
\usepackage{amsfonts}
\usepackage{graphicx}
\usepackage{bm}
\usepackage{cite}
\usepackage{times}

\date{}

\newcommand{\Section}[1]{\section{\normalsize #1}}
\newcommand{\Abstract}[0]{\section*{\normalsize Abstract}}
\newcommand{\References}[1]
 {
  \begin{thebibliography}{XX}
  \setlength{\itemsep}{-\parsep}
  #1
  \end{thebibliography}
 }
\newcommand{\Figure}[2]
 {
 \begin{figure}
  \begin{center}
 \includegraphics[width=7.5cm]{#1}
 \caption{#2}
  \end{center}
 \end{figure}
 }%
\newcommand{\Fig}[3]
 {%
 \begin{figure}
  \begin{center}
 \includegraphics[#3]{#1}
 \caption{#2}
 \end{center}
 \end{figure}%
 }%
 \newcommand{\ind}[1]{\textrm{\footnotesize #1}}

\begin{document}

\title{ \Large\bf Onset of Entanglement and Noise
Cross-Correlations in\\{}Two-Qubit System Interacting with Common
Bosonic Bath}

\author{\normalsize
Vladimir Privman, Dmitry Solenov, and Denis Tolkunov
\\\normalsize
Department of Physics, Clarkson University, Potsdam, New York 13699, USA
\\\normalsize
E-mails: Privman@clarkson.edu, Solenov@clarkson.edu, and
Tolkunov@clarkson.edu }

\maketitle \thispagestyle{empty}

\Abstract

We summarize our recent results \cite{STPs,STPb} for the induced
exchange interaction due to thermal bosonic environment (bath)
which also generates quantum noise. Our focus here is on the onset
of the interaction. We demonstrate that the induced interaction
can be used to manipulate and create entanglement over time scales
sufficiently large for controlling the two-qubit system for quantum
computing applications, though ultimately the noise effects will
dominate.

\Section{Introduction}

Recently it was demonstrated \cite{STPs,STPb,Braun,Porras} that
two qubits subject to common thermal bosonic environment (bath)
can develop considerable entanglement. A similar result has also
been obtained for qubits interacting via fermionic environment
\cite{RKKY}. Here we review our
results on the derivation of the induced exchange interaction and
quantum noise in a unified formulation \cite{STPs,STPb}, focusing
the presentation on the onset and development of the cross-qubit
correlations due to the bath.

We consider a 1D channel model for the bath, motivated by recent
experiments \cite{Experiment}, and allow bosons (e.g. phonons,
photons) to propagate along a single direction with wave vector
$k$ and dispersion $\omega_k=c_s k$. More general results are
available in \cite{STPb}. Two qubits immersed in this environment
are separated by distance $r_2-r_1=|\mathbf{d}|$ such that the
interaction due to the wave function overlap is negligible. The
qubits' interaction with the bosonic bath is introduced
\cite{STPs,Leggett} as
\begin{equation}\label{eq:S1:H_SB}
H_{SB}=\sum\limits_{j=1,2}\sigma^j_x X^j_m,
\end{equation}
where $\sigma^j_x$ is the standard Pauli matrix of qubit $j=1$ and 2, and
\begin{equation}\label{eq:S1:X_jm}
X^j_m=\sum_k g^m_k \left(a_k e^{ikr_j} + a_k^\dagger e^{-ikr_j}
\right).
\end{equation}
The total Hamiltonian is $H=H_S+H_B+H_{SB}$, where
$H_B=\sum_{k}\omega _{k}a_{k}^{\dagger}a_{k}$, $H_S$ represents
the Hamiltonian of the qubit system, and we set $\hbar=1$. The
reduced density matrix that describes the dynamics of the qubit
system is, then, given as the trace of the total density matrix
over the bath modes,
\begin{equation}\label{eq:ro_S}
\rho _S (t) = Tr_B(e^{-iHt}\rho_S(0)\rho_B e^{iHt}),
\end{equation}
where the initial density matrix is assumed factorized and
consists of the system and bath parts. The latter is
$\rho_B=e^{-H_B/kT} / Tr_B(e^{-H_B/kT})$. For large times, a more
realistic model of the environment assumes rethermalization, and
Markovian schemes are appropriate for the description of the
dynamics \cite{STPb}. However, for short times the present
formulation is adequate and provides a useful solvable model for
the case of otherwise gapless qubits, $H_S=0$, which we consider
from now on.

\Section{Exact solution to the reduced density matrix}

With the assumptions outlined above, we
utilized bosonic operator techniques\cite{STPs} to derive an exact expression
\begin{equation}\label{eq:S3:AdiabaticSolution}
\rho _S (t) = \sum\limits_{\lambda ,\lambda '} {P_\lambda  \rho _S
(0)P_{\lambda '} e^{\frak{L}_{\lambda \lambda '} (t)} }.
\end{equation}
Here the projection operator is defined as
$P_\lambda=\left|{\lambda_1\lambda_2}\right\rangle\left\langle
{\lambda _1\lambda _2}\right|$, with
$\left|{\lambda_j}\right\rangle$ the eigenvectors of
$\sigma_x^j$. The real part of the exponent in
(\ref{eq:S3:AdiabaticSolution}) leads to decay of off-diagonal
density-matrix elements resulting in decoherence,
\begin{eqnarray}\nonumber
  \textrm{Re} \frak{L}_{\lambda \lambda '}(t)&=&\!\!-\!\!\sum\limits_k {G_k (t,T)
  \left[ {\left( {\lambda '_1  - \lambda _1 } \right)^2  + \left({
  \lambda '_2  - \lambda _2 } \right)^2 } \right.}
\\ \label{eq:S3:DecoherenceFunction-ReL}
   &+&\!\!\! \left. {2\cos\! \left( {\frac{{\omega _k \left| {\mathbf{d}} \right|}}
{{c_s }}} \right)\!\left( {\lambda '_1  - \lambda _1 }
\right)\left( {\lambda '_2  - \lambda _2 } \right)} \right].
\end{eqnarray}
The imaginary part, yielding the induced interaction, is
\begin{equation}\label{eq:S3:DecoherenceFunction-ImL}
\textrm{Im} \frak{L}_{\lambda \lambda '} (t) =   \sum\limits_k
{C_k (t)\cos \left( {\frac{{\omega _k \left| {\mathbf{d}}
\right|}} {{c_s }}} \right)\left( {\lambda _1 \lambda _2  -
\lambda '_1 \lambda '_2 } \right)}.
\end{equation}
We defined the standard ``spectral''
functions \cite{Leggett,Privman}
\begin{equation}\label{eq:S3:DecoherenceFunction-Gk}
G_k (t,T) = 2\frac{{\left| {g_k } \right|^2 }} {{\omega _k^2
}}\sin ^2 \left( \frac{{\omega _k t}} {{2}}\right) \coth \left(
{\omega _k \over 2k_B T}  \right),
\end{equation}
\begin{equation}\label{eq:S3:DecoherenceFunction-Ck}
C_k (t) = 2\frac{{\left| {g_k } \right|^2 }} {{\omega _k^2
}}\left( {\omega _k t - \sin \omega _k t} \right).
\end{equation}
To evaluate (\ref{eq:S3:DecoherenceFunction-ReL}) and
(\ref{eq:S3:DecoherenceFunction-ImL}), we consider the model in
which the density of modes together with the coupling constants
are approximated by the power-law function of the frequency with
superimposed exponential cutoff \cite{Leggett}, i.e.,
\begin{equation}\label{eq:Dg_w}
\sum_k |g_k|^2 \rightarrow \alpha_n\int_0^\infty \! d\omega \, \omega^n
\exp(-\omega/\omega_c).
\end{equation}
For $n=1$ this corresponds to the well known Ohmic model
\cite{Leggett}.

\Section{Induced cross-qubit interaction and noise}

One can show that if the real part of $\frak{L}_{\lambda \lambda
'}(t)$ were absent, the exponential involving the imaginary part would
yield coherent dynamics with the unitary evolution operator
$\exp[-i\left(H_\ind{int}+F(t)\right)t]$. The constant Hamiltonian
$H_\ind{int}$ represents the induced interaction,
\begin{equation}\label{eq:S3:H-int}
H_\ind{int}\! = \! - \frac{2\alpha _n \Gamma (n)c_s^n\omega _c^n }
{\left(c_s^2 + \omega _c^2 \left| {\mathbf{d}} \right|^2
\right)^{n/2} } \cos \left[{
 n\arctan \left( {\frac{{\omega _c \left| {\mathbf{d}} \right|}}
{{c_s }}} \right)} \right]\!\sigma _x^1 \sigma _x^2.
\end{equation}
The time dependent term is given by
\begin{equation}\label{eq:S3:F(t)}
F(t) = 2\sigma _x^1 \sigma _x^2 \alpha_n\int\limits_0^\infty
{d\omega \omega^{n-1} e^{-\frac{\omega}{\omega_c}} \frac{{\sin
\omega t}} {{\omega t}}\cos \left( \frac{\omega |\mathbf{d}|}{c_s}
\right)} ;
\end{equation}
$F(t)$ commutes with $H_\ind{int}$ and therefore could be viewed
as the initial time-dependent modification of the interaction
during its onset: $F(t)$ vanishes for large times as $\alpha_n
\omega_c^n/(\omega_c t)^n$, but note that $F(0)=-H_\ind{int}$.

The interaction Hamiltonian (\ref{eq:S3:H-int}) is consistent with
the results obtained \cite{STPb} within a perturbative Markovian
approach, for more general cases. In Figure~1, we plot the
magnitude of the interaction Hamiltonian ${\cal H}_\ind{int}$,
defined via $H_\ind{int}={\cal H}_\ind{int}\sigma _x^1 \sigma
_x^2$, as a function of the qubit-qubit separation for various
$n$. At large distances the interaction decreases as
$\left|{\mathbf{d}}\right|^{-n}$, for even $n$, and
$\left|{\mathbf{d}}\right|^{-n-1}$, for odd $n$. This means, for
instance, that for spins (as qubits) with $n=1,2$, the induced
interaction decreases slower as compared to the dipole-dipole
magnetic interaction; see estimates for semiconductor impurity
electron spins in \cite{STPb}.
\Figure{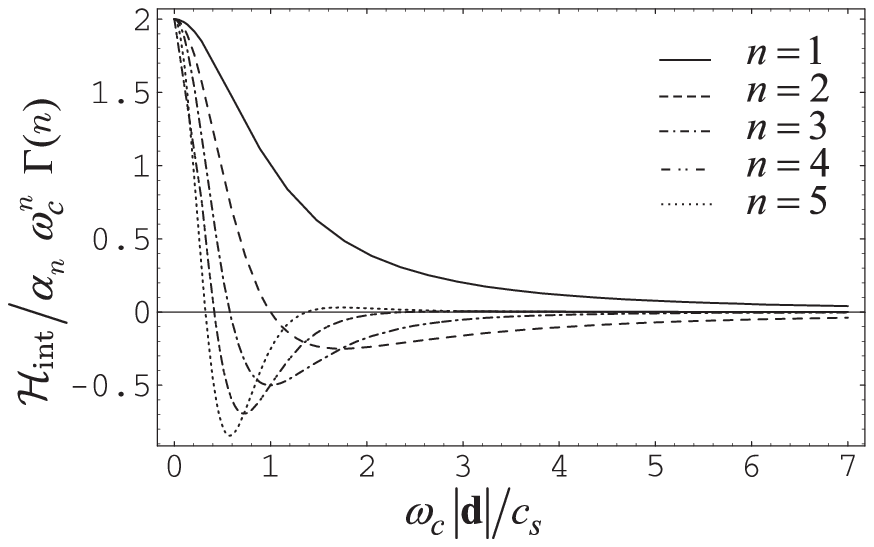}{The magnitude of the induced interaction for the
Ohmic ($n=1$) and super-Ohmic ($n>1$) bath models as a function of
qubits' separation.}

The decoherence terms, (\ref{eq:S3:DecoherenceFunction-ReL}),
describe quantum noise that ultimately destroys the coherent
dynamics given by $H_\ind{int}$ (and $F(t)$). To study the effect
of these terms, we evaluate the concurrence \cite{Wootters1} which
measures the entanglement of the spin system and is monotonically
related to the entanglement of formation \cite{Bennett-Vedral}.
For a mixed state of two qubits we first define the spin-flipped
state, $ \tilde \rho _S = \sigma^1_y \sigma^2_y \, \rho^*_S \,
\sigma^1_y \sigma^2_y $, and then the Hermitian operator
$R=\sqrt{\sqrt{\rho_S}\tilde\rho_S\sqrt{\rho_S}}$, with
eigenvalues $\lambda_{i=1,2,3,4}$. The concurrence is then given
\cite{Wootters1} by
\begin{equation}\label{eq:S3:concurence}
C\left( {\rho _S } \right) = \max \{ {0,2\max \limits_i \lambda _i
- \lambda _1 - \lambda _2 - \lambda _3 - \lambda _4} \} .
\end{equation}
\Figure{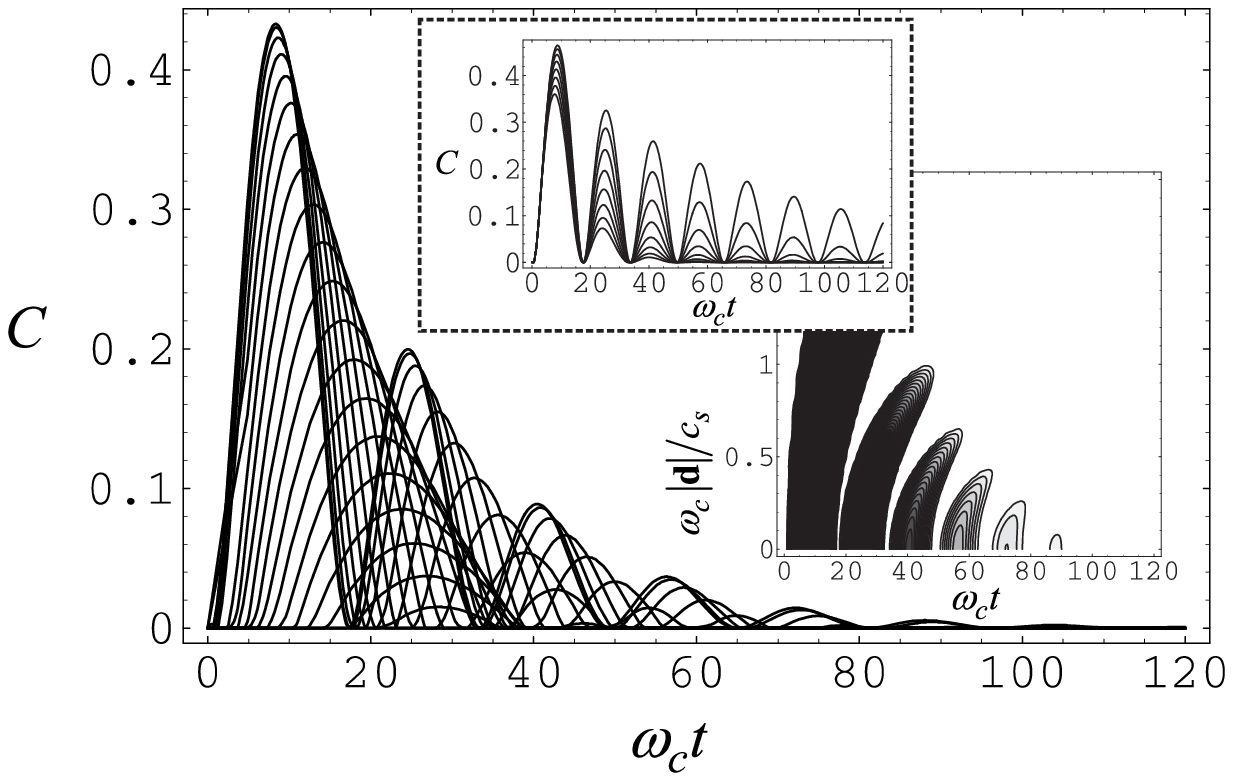}{Concurrence as a function of time for various
distances between the qubits. The right-bottom inset demonstrates
the topology of the concurrence in distance-time plane. The
parameters are $\alpha_1=1/20$, $k_BT/\omega_c=1/20$, $n=1$. The
top inset shows dynamics for different temperatures: $80
k_BT/\omega_c=1,2,3,4,5,6,7,8$.}

In Figure 2, we plot the concurrence as a function of time and the
qubit-qubit separation, for the (initially unentangled) state
$\left|\uparrow\uparrow\right\rangle$, and $n=1$. The
bath-mediated interaction between the qubits creates entanglement,
which oscillates according to the magnitude of $H_\ind{int}$. The
same bath also damps the oscillations destroying the entanglement
for larger times. The decay rate of the envelope is proportional
to the temperature, as shown in the inset of Figure~2. For the
corresponding dynamics of the density matrix elements see Section~5.

\Section{Onset of the interaction term}

Let us now investigate in greater detail the onset of the induced
interaction the time-dependence of which is given by $F(t)$. In
Figure~3, we plot the magnitude defined via $F(t)={\cal F}(t)
\sigma _x^1 \sigma _x^2$, as a function of time for various
qubit-qubit separations and $n=1$. The correction is initially
non-monotonic, but decreases for larger times as mentioned above.
The behavior for other non-Ohmic regimes is initially more
complicated, however the large time behavior is similar.

It may be instructive to consider the time dependent correction,
$H_F(t)$, to the interaction Hamiltonian during the initial
evolution, corresponding to $F(t)$. Since $F(t)$ commutes with
itself at different times, it generates unitary evolution
according to $\exp[-i\int_0^t dt' H_F(t')]$, with
$H_F(t)=d[tF(t)]/dt$,
\begin{eqnarray}\label{eq:H_F}
H_F(t)&=&\sigma _x^1 \sigma _x^2 \alpha_n\Gamma(n)
\\\nonumber
&\times&[%
u(\omega_c|\mathbf{d}|/c_s - w_c t)%
+u(\omega_c|\mathbf{d}|/c_s + w_c t)%
],
\end{eqnarray}
where $u(\xi)=\cos[n\arctan(\xi)]/[1+\xi^2]^{n/2}$. The above
expression is a superposition of two waves propagating in opposite
directions. In the Ohmic case, $n=1$, the shape of the wave is
simply $u(\xi)=1/(1+\xi^2)$. In Figure~4, we present the amplitude
of $H_F(t)$, defined via $H_F(t)={\cal H}_F\sigma _x^1 \sigma _x^2$,
as well as the sum of $H_\ind{int}$ and $H_F(t)$, for $n=1$. One
can observe that the ``onset wave'' of considerable amplitude and
shape $u(\xi)$ propagates once between the qubits, ``switching on''
the interaction. It does not affect the qubits once the
interaction has set in.
\Figure{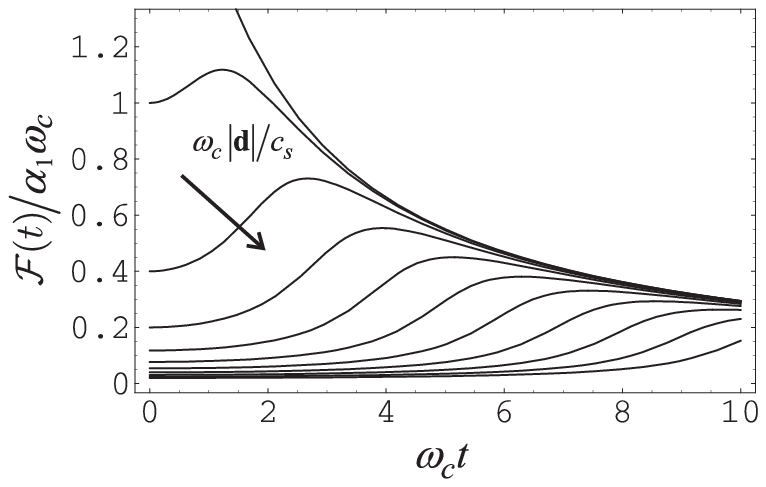}{Initial correction to the induced interaction
vs. time, for various distances between the qubits:
$\omega_c|\mathbf{d}|/c_s=0,1,\dots,10$. The Ohmic ($n=1$) case is
shown. }
\Figure{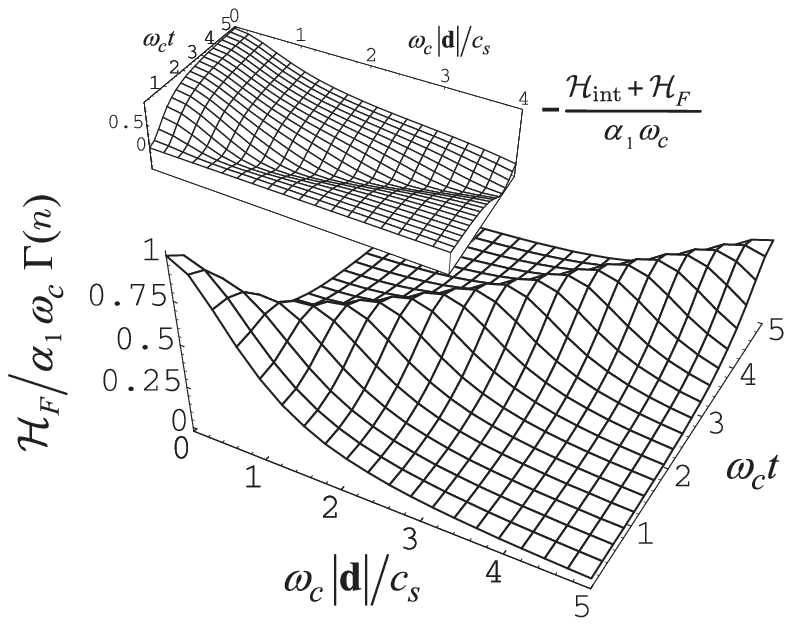}{The magnitude of the time-dependent Hamiltonian
corresponding to the initial correction as a function of time and
distance.  The Ohmic ($n=1$) case is shown. The inset demonstrates
the onset of the cross-qubit interaction on the same time scale.}

\Section{Dynamics of the density matrix elements}

To understand the dynamics of the qubit system and its
entanglement, let us again begin with the analysis of the coherent
part in (\ref{eq:S3:AdiabaticSolution}). After the interaction,
$H_\ind{int}$, has set in, it will split the system energies into
two degenerate pairs, $E_0 = E_1=-{\cal H}_\ind{int}$ and $E_2 =
E_3={\cal H}_\ind{int}$. The wave function is then
$|\psi(t)\rangle=\exp[-iH_\ind{int}t]|\psi(0)\rangle$. For the
initial ``up-up'' state,
$|\psi(0)\rangle=\left|\uparrow\uparrow\right\rangle$, it develops
as $\left|\psi(t)\right\rangle=
\left|\uparrow\uparrow\right\rangle \cos {\cal H}_\ind{int}t +
\left|\downarrow\downarrow\right\rangle i\sin {\cal
H}_\ind{int}t$. One can easily notice that at times
$t_E=\pi/4{\cal H}_\ind{int},3\pi/4{\cal H}_\ind{int},\ldots$,
maximally entangled states are obtained, while at times
$t_0=0,\pi/2{\cal H}_\ind{int},\pi/{\cal H}_\ind{int},\ldots$, the
entanglement vanishes; these special times can also be seen in Figure~2.

The bath also induces decoherence that
enters via (\ref{eq:S3:DecoherenceFunction-ReL}). The result for
the entanglement is that the decaying envelope function is
superimposed onto the coherent dynamics described above. The
magnitudes of the first and subsequent peaks of the concurrence
are determined only by this function. As temperature increases, the envelope
decays faster resulting in lower values of the concurrence, see
the inset in Figure~2.

Note also that generation of entanglement is possible only
provided that the initial state is a superposition of the
eigenvectors of the induced interaction with more than one eigenvalue
(for pure initial states). For example, the initial state
$(\left|\downarrow\uparrow\right\rangle +
\left|\uparrow\downarrow\right\rangle)/\sqrt{2}$ in our case would
only lead to the destruction of entanglement, i.e., monotonically
decreasing concurrence, similar to the results in
\cite{Eberly1}.

Since $H_S=0$, there is no relaxation by energy transfer in the system, and the
exponentials in (\ref{eq:S3:AdiabaticSolution}), with
(\ref{eq:S3:DecoherenceFunction-ReL}), suppress only the
off-diagonal matrix elements, i.e., those with
$\lambda\neq\lambda'$. It happens, however that at large times the
$\mathbf{d}$-dependence is not important in
(\ref{eq:S3:DecoherenceFunction-ReL}) and
$\textrm{Re}\frak{L}_{\lambda\lambda'}(t\rightarrow\infty)$
vanishes for certain values of $\lambda\neq\lambda'$. In the
basis of the qubit-bath interaction, $\sigma_x^1\sigma_x^2$, the
limiting $t\to \infty$ density matrix for our initial state ($\left|\uparrow\uparrow\right\rangle$) is
$\frac{1}{4}+\frac{1}{4}\left|+-\right\rangle\left\langle-+\right|
+\frac{1}{4}\left|-+\right\rangle\left\langle+-\right|$, which
takes the form
\begin{equation}\label{eq:roLimit}
\rho(t\rightarrow\infty)\rightarrow\frac{1} {8}\left(
{ {\begin{array}{*{20}c}
   3 & 0 & 0 & { - 1}  \\
   0 & 1 & 1 & 0  \\
   0 & 1 & 1 & 0  \\
   { - 1} & 0 & 0 & 3  \\
 \end{array} }} \right)
\end{equation}
in the basis of states $\left|\uparrow\uparrow\right\rangle$,
$\left|\uparrow\downarrow\right\rangle$,
$\left|\downarrow\uparrow\right\rangle$, and
$\left|\downarrow\downarrow\right\rangle$. The significance of such results, see also \cite{Braun},
is that in the present model not all the off-diagonal matrix elements are suppressed by decoherence, even though the concurrence of this mixed state is zero.

The probabilities for the qubits to occupy the states
$\left|\uparrow\uparrow\right\rangle$,
$\left|\uparrow\downarrow\right\rangle$,
$\left|\downarrow\uparrow\right\rangle$, and
$\left|\downarrow\downarrow\right\rangle$ are presented in Figure~5.
For the initial state $\left|\uparrow\uparrow\right\rangle$,
only the diagonal and inverse-diagonal matrix elements are
affected, and the system oscillates between the two states
$\left|\uparrow\uparrow\right\rangle$ and
$\left|\downarrow\downarrow\right\rangle$, as mentioned earlier in
the description of the coherent dynamics, while decoherence dampens
these oscillations down. In addition, decoherence actually raises the other two diagonal
elements to a certain level, see (\ref{eq:roLimit}). The
dynamics of the inverse-diagonal density matrix elements is shown
in Figure~6.
\Fig{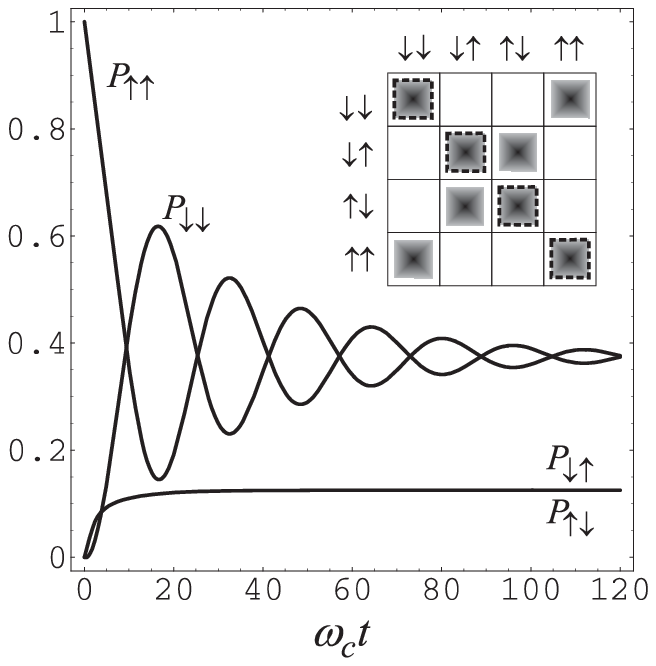}{Dynamics of the occupation probabilities for the states
$\left|\uparrow\uparrow\right\rangle$,
$\left|\uparrow\downarrow\right\rangle$,
$\left|\downarrow\uparrow\right\rangle$, and
$\left|\downarrow\downarrow\right\rangle$. The parameters are the
same as in Figure~1. The inset shows the structure of the reduced
density matrix (the non-shaded entries are zeros).}{width=5cm}
\Fig{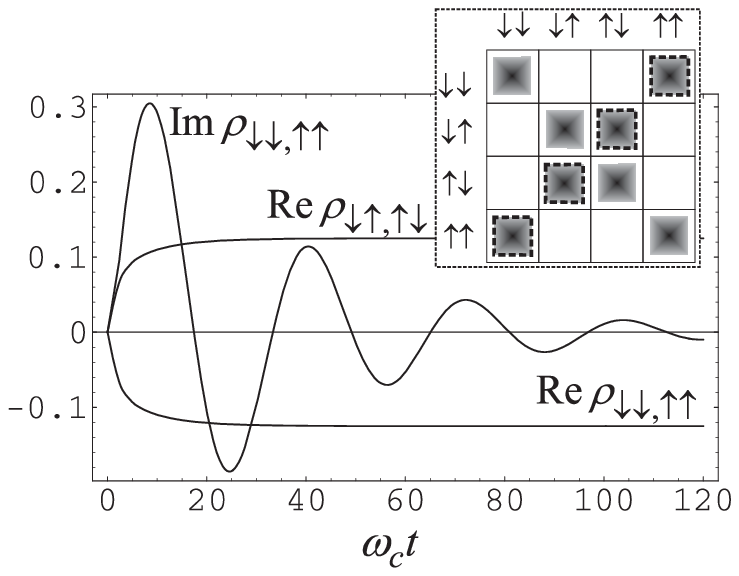}{Dynamics of the off-diagonal matrix elements for the same system as in Figure~5.}{width=5.6cm}

\Section{Conclusion}

To summarize, we studied the initial stages of the
cross-qubit interaction induced by a thermal bosonic bath. It was shown that thermal environment can create a sufficiently
large entanglement for quantum control, though it is erased for larger times. The dynamics of
the entanglement and the density matrix elements have been
investigated.

This research was supported by the NSF under grant
DMR-0121146.

\References {\frenchspacing

\bibitem{STPs} D. Solenov, D. Tolkunov, and V. Privman, Phys. Lett. A (in print), cond-mat/0511680.

\bibitem{STPb} D. Solenov, D. Tolkunov, and V. Privman, cond-mat/0605278.

\bibitem{Braun} D. Braun, Phys. Rev. Lett. \textbf{89}, 277901 (2002).

\bibitem{Porras} D. Porras and J. I. Cirac, Phys. Rev. Lett. \textbf{92}, 207901 (2004).

\bibitem{RKKY}  M. A. Ruderman and C. Kittel, Phys. Rev. \textbf{96}, 99 (1954); T. Kasuya, Prog. Theor. Phys. \textbf{16}, 45 (1956);
K. Yosida, Phys. Rev. \textbf{106}, 893 (1957);
V. Privman, I. D. Vagner, and G. Kventsel, Phys. Lett. A \textbf{239}, 141 (1998); C. Piermarocchi, P. Chen, L. J. Sham, and D. G. Steel, Phys. Rev. Lett. \textbf{89}, 167402 (2002); D. Mozyrsky, A. Dementsov, and V. Privman, Phys. Rev. B \textbf{72}, 233103 (2005); Y. Rikitake and H. Imamura, Phys. Rev. B \textbf{72}, 033308 (2005).

\bibitem{Experiment} N. J. Craig, J. M. Taylor, E. A. Lester, C. M. Marcus, M.
P. Hanson, and A. C. Gossard, Science \textbf{304}, 565 (2004); J.
M. Elzerman, R. Hanson, L. H. Willems van Beveren, B. Witkamp, L.
M. K. Vandersypen, and L. P. Kouwenhoven, Nature \textbf{430}, 431
(2004); M. R. Sakra, H. W. Jiang, E. Yablonovitch, and E. T. Croke,
Appl. Phys. Lett. \textbf{87}, 223104 (2005).

\bibitem{Leggett} A. J. Leggett, S. Chakravarty, A. T. Dorsey, M. P. A. Fisher, A.
Garg, and W. Zwerger, Rev. Mod. Phys. \textbf{59}, 1 (1987).

\bibitem{Privman}  V. Privman, Modern Phys. Lett. B \textbf{16}, 459 (2002).

\bibitem{Wootters1} S. Hill and W. K. Wootters, Phys. Rev. Lett. \textbf{78}, 5022 (1997); W. K. Wootters, Phys. Rev. Lett. \textbf{80}, 2245 (1998).

\bibitem{Bennett-Vedral} C. H. Bennett, D. P. DiVincenzo, J. Smolin, and W. K. Wootters, Phys. Rev. A \textbf{54}, 3824 (1996).

\bibitem{Eberly1} T. Yu and J. H. Eberly, Phys. Rev. B \textbf{68}, 165322 (2003); T. Yu and J. H. Eberly, Phys. Rev. Lett. \textbf{93}, 140404 (2004).

}

\end{document}